# Effects of a mesoporous bioactive glass on osteoblasts, osteoclasts and macrophages

N. Gómez-Cerezo[a,b], L. Casarrubios[c], I. Morales[c], M. J. Feito[c], M. Vallet-Regí[a,b*],

D. Arcos[a,b*], M. T. Portolés[c*]

[a] *Departamento de Química en Ciencias Farmacéuticas, Facultad de Farmacia, Universidad Complutense de Madrid, Instituto de Investigación Sanitaria Hospital 12 de Octubre i+12, Plaza Ramón y Cajal s/n, 28040 Madrid, Spain.*

[b] *CIBER de Bioingeniería, Biomateriales y Nanomedicina, CIBER-BBN, Madrid, Spain.*

[c] *Departamento de Bioquímica y Biología Molecular, Facultad de Ciencias Químicas, Universidad Complutense de Madrid, Instituto de Investigación Sanitaria del Hospital Clínico San Carlos (IdISSC), 28040-Madrid, Spain.*

\* Corresponding authors

*E-mail address:* portoles@quim.ucm.es, arcosd@ucm.es, vallet@ucm.es

# Abstract

A mesoporous bioactive glass (MBG) of molar composition $75SiO_2$-$20CaO$-$5P_2O_5$ (MBG-75S) has been synthetized as a potential bioceramic for bone regeneration purposes. X-ray diffraction (XRD), Fourier transform infrared spectroscopy (FT-IR), nitrogen adsorption studies and transmission electron microscopy (TEM) demonstrated that MBG-75S possess a highly ordered mesoporous structure with high surface area and porosity, which would explain the high ionic exchange rate (mainly calcium and silicon soluble species) with the surrounded media. MBG-75S showed high biocompatibility in contact with Saos-2 osteoblast-like cells. Concentrations up to 1 mg/ml did not lead to significant alterations on either morphology or cell cycle. Regarding the effects on osteoclasts, MBG-75S allowed the differentiation of RAW-264.7 macrophages into osteoclast-like cells but exhibiting a decreased resorptive activity. These results point out that MBG-75S does not inhibit osteoclastogenesis but reduces the osteoclast bone-resorbing capability. Finally, *in vitro* studies focused on the innate immune response, evidenced that MBG-75S allows the proliferation of macrophages without inducing their polarization towards the M1 pro-inflammatory phenotype. This *in vitro* behavior is indicative that MBG-75S would just induce the required innate immune response without further inflammatory complications under *in vivo* conditions. The overall behavior respect to osteoblasts, osteoclasts and macrophages, makes this MBG a very interesting candidate for bone grafting applications in osteoporotic patients.

# 1. Introduction

Mesoporous bioactive glasses (MBGs) are bioceramics intended for bone tissue regeneration purposes. Discovered in 2004 by Zhao *et al* [1], MBGs mean a significant upgrade respect to the conventional sol-gel bioactive glasses prepared by Li *et al* in 1991 [2]. Similarly, to sol-gel bioactive glasses, MBGs are commonly prepared in the ternary system $SiO_2$-CaO-$P_2O_5$ [3-5] and, in the last decade, different research groups have incorporated different ions with potential therapeutic properties [6-11]. In the case of MBGs, the incorporation of a structure directing agent (SDA) to the synthesis results in the formation of an ordered mesophase by the self-organization of the SDA into micelles. Soluble silica, phosphate and calcium species condensates around this organic template, which leads to a mesoporous structure after calcination, thus providing higher textural properties compared to conventional sol-gel bioactive glasses [12,13].

The primary consequence on the biological behavior is a faster and more intense ionic exchange (mainly $Ca^{2+}$ and silica species) between the MBG and the surrounding fluids [14]. In fact, some MBGs have shown the fastest *in vitro* bioactive behavior when soaked in simulated body fluid, in terms of the nucleation and growth of a carbonate nanocrystalline apatite on their surface, very similar to the biological one found in bones [15]. However, the MBG surface reactivity is not their only action mechanism. The ions released from MBG also stimulate the expression of several genes of osteoblastic cells and induce angiogenesis both *in vitro* and *in vivo* [16,17]. Recent studies suggest that these ions could also regulate immune responses by altering the ionic microenvironment between the implants and hosts [18]. The importance of the immune response during biomaterial-mediated osteogenesis makes necessary the evaluation of the osteoimmunomodulatory properties of biomaterials for bone tissue [19].

Recently, the *in vivo* response to these materials has been studied in different animal models, evidencing certain advantages respect to other bioceramics. Due to their potential bone regeneration capabilities, MBGs are being considered as bone grafts in the case of osteoporotic patients. Osteoporosis is produced by the bone remodeling disruption that is due to either increased bone resorption by osteoclasts or decreased new bone formation by osteoblasts or both [20]. The biomaterials most commonly employed for treatment of osteoporotic bone and bone regeneration have been designed to stimulate the osteogenesis process and bone formation by osteoblasts. For this reason, osteoblasts are commonly used for the *in vitro* evaluation of bone materials [21] but few studies are focused on the effects of these biomaterials on bone resorbing osteoclasts [22]. Osteoclasts are multinucleated giant cells which differentiate from hematopoietic stem cells of the monocyte/macrophage lineage through sequential steps [23] regulated by several growth factors and cytokines expressed by different bone cell types [24,25]. Osteoclasts can also differentiate *in vitro* from macrophages by stimulation with the macrophage/monocyte-colony-stimulating factor (M-CSF) and the receptor activator of nuclear factor kappa-B ligand (RANKL) [22]. These agents induce the fusion of pre-osteoclasts, which become multinucleated cells, and the formation of "ruffled membrane", critical for bone resorption, that involves the tight attachment of osteoclasts to the bone surface to create the "sealing zone" rich in F-actin [26]. During bone resorption, osteoclasts isolate the resorptive space from the surrounding bone and release matrix-degrading enzymes, hydrogen ions and chloride ions inside the sealing zone, producing the bone matrix degradation and the dissolution of the bone mineral component, respectively [27].

The effects of MBGs on osteoblasts have been widely evaluated by different research groups [7,12,13,39,42], whereas the effects on other cell types involved in bone remodeling are practically unknown. The present study is focused on the effects of a potential mesoporous bioactive glass for bone regeneration, with molar composition $75SiO_2$-$20CaO$-$5P_2O_5$ (MBG-75S), on osteoclast differentiation, bone resorption activity and macrophage activation towards pro-inflammatory M1 phenotype. CaO plays a fundamental role in the biological properties of $SiO_2$-$CaO$-$P_2O_5$ MBGs. However, previous works demonstrated that compositions with higher CaO content led to disordered mesoporous structures [4]. The composition MBG-75S was chosen with the aim of ensuring enough CaO content while keeping the highly ordered mesoporous structure. Previously, the dose-dependent action of this powdered material on osteoblasts has been evaluated through the analysis of cell cycle, morphology, size, complexity and apoptosis after the treatment with different doses of MBG-75S.

## 2. Materials and methods

*2.1. Synthesis and characterization of MBG-75S*

Mesoporous bioactive glass MBG-75S with molar composition $75SiO_2$-$20CaO$-$5P_2O_5$ was prepared by EISA method and using Pluronic F127 as structure directing agent. For this purpose, 32 g of Pluronic F127 was dissolved in an ethanol-HCl (0.5M) solution. Thereafter, 61.3 ml of tetraethylorthosilane (TEOS), 6.28 ml of triethylphosphate (TEP) and 17.6 mg of $Ca(NO_3)_2 \cdot 4H_2O$ were gradually added in 3 hours intervals. The mixture was stirred for 24 hours, poured into Petri dishes (9 cm in diameter) and introduced in an incubator at 30ºC for 7 days, until solvent evaporation and gelling. The transparent membranes so obtained were calcined at 700ºC for 3 hours under air atmosphere. The resultant powder was gently milled in dry conditions and sieved, collecting the grain fraction below 40 micrometers. Chemicals of highest purity available have been used in the present study.

X-ray diffraction experiment was carried out in a Philips X'Pert diffractometer equipped with a Cu K$\alpha$ radiation (wavelength 1.5406 Å). The patterns were collected between 0.5 and 6.5 $2\theta^o$ angle using a Bragg-Brentano geometry. Fourier-transform infrared spectroscopy was done using a Nicolet Magma IR 550 spectrometer and using the attenuated total reflectance (ATR) sampling technique with a Golden Gate accessory.

Nitrogen adsorption/desorption isotherm was obtained with an ASAP 2020 equipment. The MBG-75S was previously degassed under vacuum for 15 h, at 150 ºC. The surface area was determined using the Brunauer-Emmett-Teller (BET) method. The pore size distribution between 0.5 and 40 nm was determined from the adsorption branch of the isotherm by means of the Barret-Joyner-Halenda (BJH) method. The surface area was calculated by the BET method and the pore size distribution was determined by the BJH method using the adsorption branch of the isotherm.

Scanning electron microscopy (SEM) was carried out using a JEOL-6335F microscope, operating at 15 kV. Transmission electron microscopy (TEM) was carried out using a JEOL-1400 microscope, operating at 300 kV (Cs 0.6mm, resolution 1.7 Å). Images were recorded using a CCD camera (model Keen view, SIS analyses size 1024 X 1024, pixel size 23.5mm X 23.5mm) at 60000X magnification using a low-dose condition.

*2.2. Soluble species release from MBG-75S to the culture medium*

The levels of soluble calcium, phosphates and silica species in the culture medium were measured by inductively coupled plasma (ICP) spectroscopy, after soaking MBG-75S in

Dulbecco´s Modified Eagle´s Medium (DMEM, Sigma Chemical Company, St. Louis, MO, USA) (1 mg/ml) for 3 and 7 days.

*2.3. Culture of human Saos-2 osteoblasts*

Human Saos-2 osteoblasts ($10^5$ cells/ml) were cultured in the presence of 0.5, 1 and 2 mg/ml of powdered MBG-75S in Dulbecco´s Modified Eagle´s Medium (DMEM, Sigma Chemical Company, St. Louis, MO, USA) supplemented with 10% (vol/vol) fetal bovine serum (FBS, Gibco, BRL), 1 mM L-glutamine (BioWhittaker Europe, Belgium), penicillin (200 μg/ml, BioWhittaker Europe, Belgium), and streptomycin (200 μg/ml, BioWhittaker Europe, Belgium), under a 5% $CO_2$ atmosphere and at 37ºC. Controls in the absence of material were carried out in parallel. After 24 hours, the culture medium was aspirated, the cells were washed with phosphate-buffered saline (PBS) and harvested using 0.25% trypsin-ethylene diamine tetraacetic acid (EDTA). Cell suspensions were centrifuged at 310 x g for 10 min and resuspended in fresh medium for the analysis of cell cycle, apoptosis, cell size and complexity by flow cytometry as described below.

*2.4. Cell-cycle and apoptosis analysis by flow cytometry*

Cells were resuspended in PBS (0.5 ml) and incubated with 4.5 ml of ethanol 70% during 4 hours at 4ºC. Then, cells were centrifuged at 310 x g for 10 min, washed with PBS and resuspended in 0.5 ml of PBS with Tritón X-100 0.1%, propidium iodide (IP) 20 μg/ml and ribonuclease (RNAse) 0.2 mg/ml (Sigma-Aldrich, St. Louis, MO, USA). After incubation at 37ºC for 30 min, the fluorescence of PI was excited by a 15 mW laser tunning to 488 nm and the emitted fluorescence was measured with a 585/42 band pass filter in a FACScan Becton Dickinson flow cytometer. The cell percentage in each cycle phase: $G_0/G_1$, S and $G_2/M$ was calculated with the CellQuest Program of Becton

Dickinson and the SubG$_1$ fraction (cells with fragmented DNA) was used as indicative of apoptosis. For statistical significance, at least 10,000 cells were analyzed in each sample.

*2.5. Cell size and complexity detection by flow cytometry*

Forward angle (FSC) and side angle (SSC) scatters were evaluated as indicative of cell size and complexity, respectively, using a FACScan Becton Dickinson flow cytometer.

*2.6. Osteoclast differentiation from murine RAW 264.7 macrophages*

Murine RAW-264.7 macrophages (2 x 10$^4$ cells/ml) were seeded on glass coverslips and cultured in the presence of 1 mg/ml of powdered MBG-75S in Dulbecco's Modified Eagle Medium (DMEM) without phenol red, supplemented with 10% fetal bovine serum (FBS, Gibco, BRL), 1 mM L-glutamine (BioWhittaker Europe, Belgium), penicillin (200 μg/ml, BioWhittaker Europe, Belgium), and streptomycin (200 μg/ml, BioWhittaker Europe, Belgium). To stimulate osteoclast differentiation, 40 ng/ml of mouse receptor activator of nuclear factor kappa-B ligand (RANKL) recombinant protein (TRANCE/RANKL, carrier-free, BioLegend, San Diego) and 25 ng/ml recombinant human macrophage-colony stimulating factor (M-CSF, Milipore, Temecula) were added to the culture medium. Cells were cultured under a 5% CO$_2$ atmosphere and at 37ºC for 7 days. Controls in the absence of material were carried out in parallel.

*2.7. Lactate dehydrogenase (LDH) measurement*

To evaluate the plasma membrane integrity during osteoclast differentiation, lactate dehydrogenase (LDH) activity was measured in the culture medium by an enzymatic method at 340 nm (Bio-Analítica) using a Beckman DU 640 UV-Visible spectrophotometer.

*2.8. Morphological studies by confocal microscopy*

Cells were seeded on glass coverslips and cultured in the presence of different doses of MBG-75S for 24 hours. Controls in the absence of material were carried out in parallel. After washing with PBS, cells were fixed with 3.7% paraformaldehyde in PBS for 10 min, permeabilizated with 0.1% Triton X-100 for 3 min and preincubated with PBS containing 1% bovine serum albumin (BSA) for 30 min. Then, cells were incubated with rhodamine phalloidin (1:40, v/v Molecular Probes) for 20 min to stain F-actin filaments. Samples were then washed with PBS and cell nuclei were stained with 3 μM 4′-6-diamidino-2′-phenylindole (DAPI, Molecular Probes) for 5 min. After staining and washing with PBS, cells were examined using a Leica SP2 Confocal Laser Scanning Microscope. Rhodamine fluorescence was excited at 540 nm and measured at 565 nm. DAPI fluorescence was excited at 405 nm and measured at 420–480 nm.

*2.9. Osteoclast resorption activity*

To evaluate the resorption activity of osteoclasts, RAW-264.7 macrophages were seeded on the surface of nanocrystalline hydroxyapatite disks and differentiate into osteoclasts in the presence of 1 mg/ml of powdered MBG-75S as it is described above. Nanocrystalline hydroxyapatite disks were prepared by controlled precipitation of calcium and phosphate salts and subsequently heated at temperatures below the sintering point, as previously described by our research group [22]. Controls in the absence of material were carried out in parallel. After 7 days of differentiation, cells were detached using cell scrapers and disks were dehydrated, coated with gold-palladium and examined with a JEOL JSM-6400 scanning electron microscope in order to observe the geometry of resorption cavities produced by osteoclasts on the surface of nano-HA disks.

*2.10. Detection of pro-inflammatory M1 macrophage phenotype*

To study the effect of MBG-75S on macrophage polarization towards pro-inflammatory M1 phenotype, RAW-264.7 macrophages were cultured with 1 mg/ml of this material for 24 hours in the presence or the absence of *E. coli* lipopolysaccharide/interferon-γ (250 ng/ml LPS and 100 ng/ml IFN, Sigma-Aldrich Corporation, St. Louis, MO, USA) as inflammatory stimuli [28] in Dulbecco's Modified Eagle Medium (DMEM) supplemented with 10% fetal bovine serum (FBS, Gibco, BRL), 1 mM L-glutamine (BioWhittaker Europe, Belgium), penicillin (200 μg/ml, BioWhittaker Europe, Belgium), and streptomycin (200 μg/ml, BioWhittaker Europe, Belgium) at 37 ºC under a $CO_2$ (5%) atmosphere. Controls in the absence of material were carried out in parallel. For the analysis of macrophage proliferation, the attached RAW-264.7 cells were washed with phosphate buffered saline (PBS), harvested using cell scrapers and counted with a Neubauer hemocytometer.

The expression of CD80 as M1 marker [29] was used to detect pro-inflammatory M1 macrophages by flow cytometry and confocal microscopy. For flow cytometry studies, cells were detached, centrifuged and incubated in 45 μl of staining buffer (PBS, 2.5% FBS Gibco, BRL) with 5 μl of normal mouse serum inactivated for 15 min at 4º C in order to block the Fc receptors on the macrophage plasma membrane and to prevent non-specific binding. Then, cells were incubated with phycoerythrin (PE) conjugated anti-mouse CD80 antibody (2.5 μg/ml, BioLegend, San Diego, California) for 30 min at 4º C in the dark. Labelled cells were analyzed using a FACSCalibur flow cytometer. PE fluorescence was excited at 488 nm and measured at 585/42 nm.

For confocal microscopy studies, macrophages cultured on glass coverslips were fixed with 3.7% paraformaldehyde (Sigma-Aldrich Corporation, St. Louis, MO, USA) in PBS for 10 min, washed with PBS and permeabilized with 0.1% Triton X-100 (Sigma-Aldrich Corporation, St. Louis, MO, USA) for 3 min. The samples were then washed with PBS and preincubated with PBS containing 1% BSA (Sigma-Aldrich Corporation, St. Louis, MO, USA) for 30 min to prevent non-specific binding. Samples were incubated in 1 ml of staining buffer with phycoerythrin (PE) conjugated anti-mouse CD80 antibody (2.5 µg/ml, BioLegend, San Diego, California) for 30 min at 4ºC in the dark. Samples were then washed with PBS and the cell nuclei were stained with 3 µM DAPI (4′-6-diamidino-2′-phenylindole, Molecular Probes) for 5 min. Samples were examined using a Leica SP2 Confocal Laser Scanning Microscope. PE fluorescence was excited at 488 nm and measured at 575-675 nm. DAPI fluorescence was excited at 405 nm and measured at 420–480 nm.

*2.11. Statistics*

Data are expressed as means + standard deviations of a representative of three repetitive experiments carried out in triplicate. Statistical analysis was performed by using the Statistical Package for the Social Sciences (SPSS) version 22 software. Statistical comparisons were made by analysis of variance (ANOVA). Scheffé test was used for *post hoc* evaluations of differences among groups. In all statistical evaluations, $p < 0.05$ was considered as statistically significant.

# 3. Results and discussion

*3.1. Synthesis and characterization of MBG-75S*

The structural, chemical and textural properties of MBG-75S were determined prior to any cell culture test. Low angle XRD pattern (Figure 1 a) shows two diffraction maxima at 1.05 and 1.75 2θº that could be assigned to the (1 0) and (1 1) reflections of p6m hexagonal planar group, as previously observed for similar MBGs prepared with F127 as SDA [12]. FTIR spectrum (Figure 1 b) shows the characteristic bands at 500 and 1080 cm$^{-1}$ corresponding to Si-O-Si bending and stretching mode, respectively. The weak adsorption band observed at 590 cm$^{-1}$ corresponds to the bending vibrational modes of $PO_4^{3-}$ groups in an amorphous environment.

The isotherm obtained by nitrogen adsorption analysis (Figure 1 c) can be described as a type IV curve, characteristic of mesoporous materials. The isotherm has a type H1 hysteresis loop in the mesopore range, which is characteristic of cylindrical pores open at both ends, having necks along the pores. The dV/d log D plot (data not shown) showed a monomodal distribution centered around 5.7 nm (see Table 1). The structural parameters obtained by XRD together with nitrogen adsorption data provide valuable information about the MBG structure. Table 1 shows the structural and textural parameters and a scheme of the MBG structure is shown in Figure 1 d. For instance, from the diffraction angle 2θº of the (1 0) maxima and using the Bragg's Law

$$n\lambda = 2d\,sen\theta \quad (eq.\ 1)$$

where n is a positive integer (in our case 1) and λ is the wavelength of the incident beam (1.5406 Å), the interplanar distance for the (0 1) was calculated as 8.41 nm. Considering the hexagonal structure of the *p6m* planar group, the lattice parameter *a* of the mesoporous structure can be easily obtained (see Table 1). In a hexagonal planar structure, the lattice parameter corresponds to the distance form center of pore to center of pore, as it is shown in Figure 1 d. Using the pore size provided by the nitrogen adsorption analysis, we can also calculate the thickness of the pore walls. SEM observations (Figure 2.a) show that MBG-75S consist in particles of irregular shape, ranging in size between 10 and 40 μm. TEM study confirmed the highly ordered mesoporous structure of the MBG-75S (Figure

2.b), as well as the channel-like morphology of the pores characteristic of a *p6m* planar group.

Taken as a whole, we can describe the structure of MBG-75S as a highly porous material, with a regular arrangement of single modal pore size distribution. Considering the distance between pores and the pore size, we can conclude that the walls of MBG-75S are thinner than the pore diameters. This fact, together with the high surface area, pore volume and channel-like open morphology of the pores (deduced from the H1 hysteresis loop and TEM images) would facilitate the fast ionic dissolution and exchange, in contact with the surrounding fluids under both *in vitro* and *in vivo* conditions.

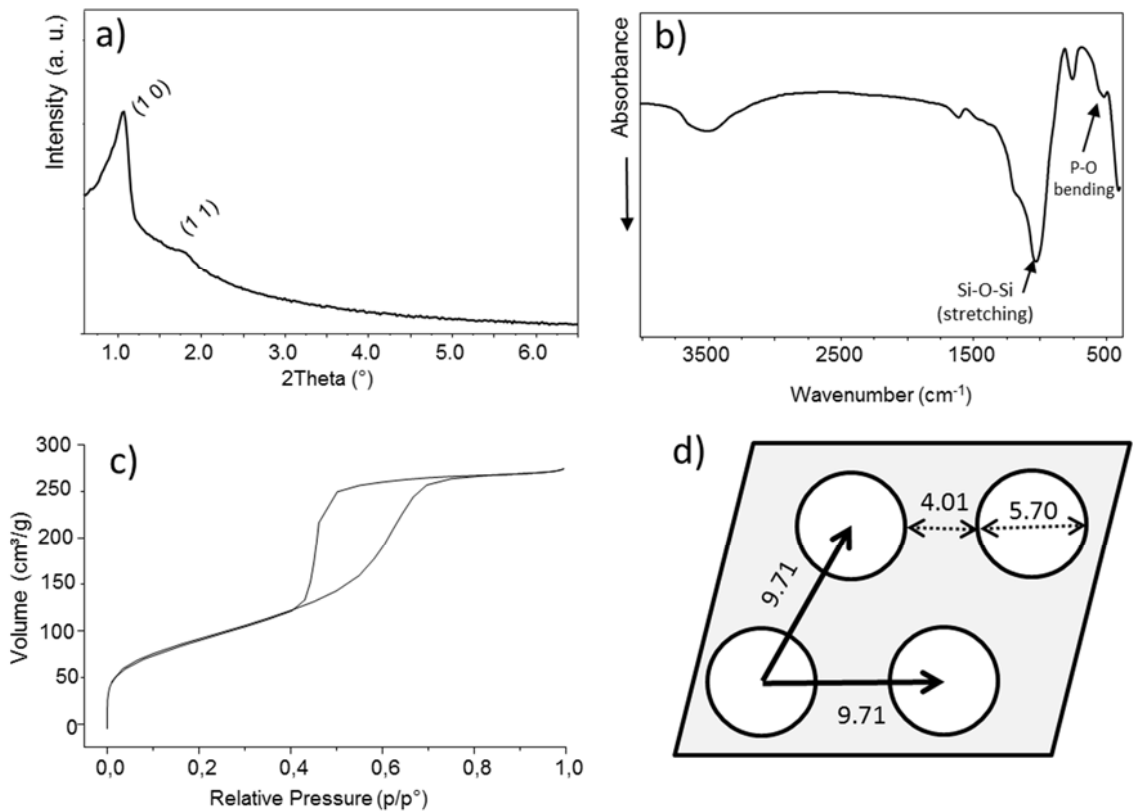

**Figure 1**. Structural, chemical and textural characterization of MBG-75S; (a) XRD pattern. The Miller indexes for a planar hexagonal unit cell p6m are indicated; (b) FTIR spectra; (c) Nitrogen adsorption/desorption isotherm and (d) a scheme of the MBG-75S porous structure calculated from the mentioned characterization techniques.

Table 1. Textural and structural parameters for MBG-S75

| Surface area (m$^2$·g$^{-1}$) | Pore Volume (cm$^3$·g$^{-1}$) | Pore size (nm) | d$_{(1\,0)}$ (nm) | Lattice parameter $a^1$ (nm) | Wall thickness$^2$ (nm) |
|---|---|---|---|---|---|
| 305.5 | 0.46 | 5.7 | 8.41 | 9.71 | 4.01 |

$^1$Calculated as $a = d_{(10)} \cdot 2/\sqrt{3}$; $^2$ Calculated as $a$ – pore size

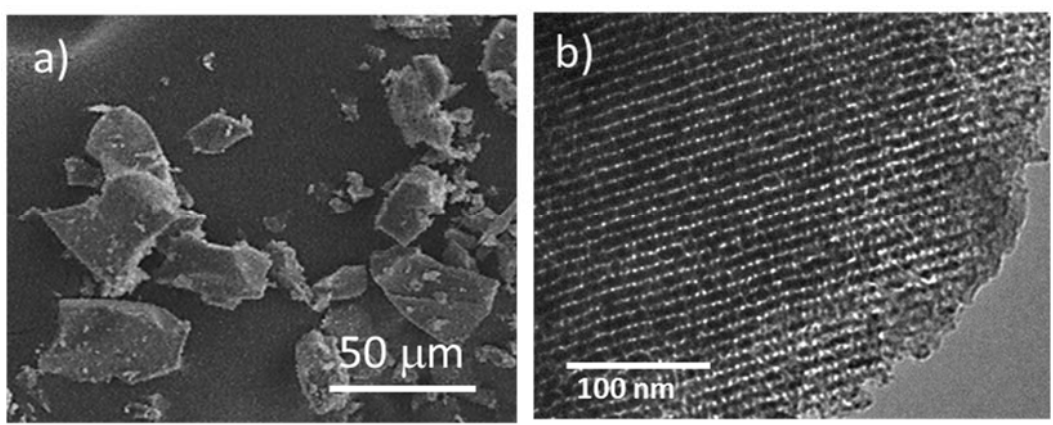

**Figure 2.** SEM micrograph (a) and TEM image (b) obtained for MBG75-S.

*3.2. Dose-dependent effects of MBG-75S on human Saos-2 osteoblasts*

The dose-dependent action of powdered MBG-75S on osteoblasts has been studied with human Saos-2 cells as *in vitro* experimental model. This osteosarcoma cell line is commonly used in this kind of *in vitro* studies due to its osteoblastic properties as production of mineralized matrix, high alkaline phosphatase levels, PTH receptors and osteonectin presence [30]. All these studies have been performed in direct contact between cells and powdered MBG. By culturing the cells in direct contact with the MBG particles, we can study not only the effects of the ions released from MBG but also the effect of the very fast release associated to the high textural properties of MBG. Different cell parameters (cell cycle, morphology, size, complexity and apoptosis) were evaluated after 24 hours of culture of Saos-2 osteoblasts with increasing doses of MBG-75S.

The analysis of the cell cycle by flow cytometry allowed us to know the effects of this MBG on the proliferation of human Saos-2 osteoblasts through progressive stages: $G_0/G_1$ phase (Quiescence/Gap1), S phase (Synthesis) and finally $G_2/M$ phase (Gap2 and Mitosis). This analysis also indicates the percentage of apoptotic cells with fragmented DNA corresponding to the $SubG_1$ fraction. Figures 3 and 4 show the cell cycle profiles of osteoblasts and the percentages of cells within each cycle phase, respectively, after 24 hours of culture in the absence or the presence of different doses of MBG-75S. As it can be observed in these figures, no alterations were detected in the $G_0/G_1$, S and the $G_2/M$ phases after treatment with 0.5 and 1 mg/ml of this MBG. However, 2 mg/ml induced significant decreases in the $G_0/G_1$ phase ($p < 0.05$) and the $G_2/M$ phase ($p<0.005$). This effect can be explained by the significant increase ($p < 0.005$) produced by 2 mg/ml in the $SubG_1$ fraction (apoptotic cells). Very low levels of apoptosis were detected either in the absence of material or in the presence of 0.5 mg/ml and 1 mg/ml of MBG-75S (lower than 5% $\pm$ 0.5%), Figure 4). The high apoptosis levels (20% $\pm$ 1%) observed with 2 mg/ml are probably due to the presence of this material in powdered form which can produce the loss of cell anchorage, inducing a kind of apoptosis defined as anoikis [31,32].

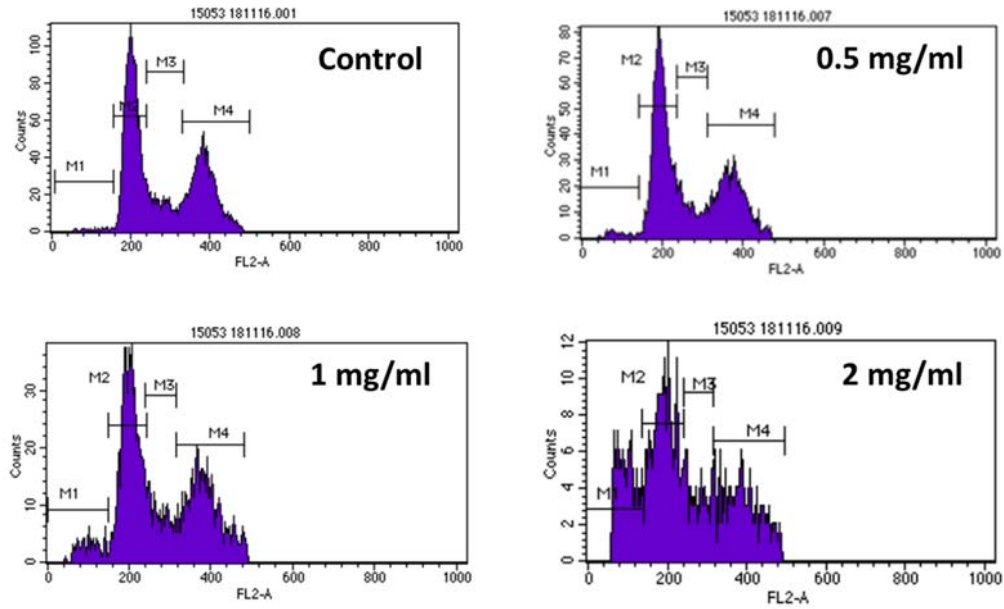

**Figure 3.** Effects of different doses of MBG-75S on cell cycle profile of human Saos-2 osteoblasts after 24 hours of treatment.

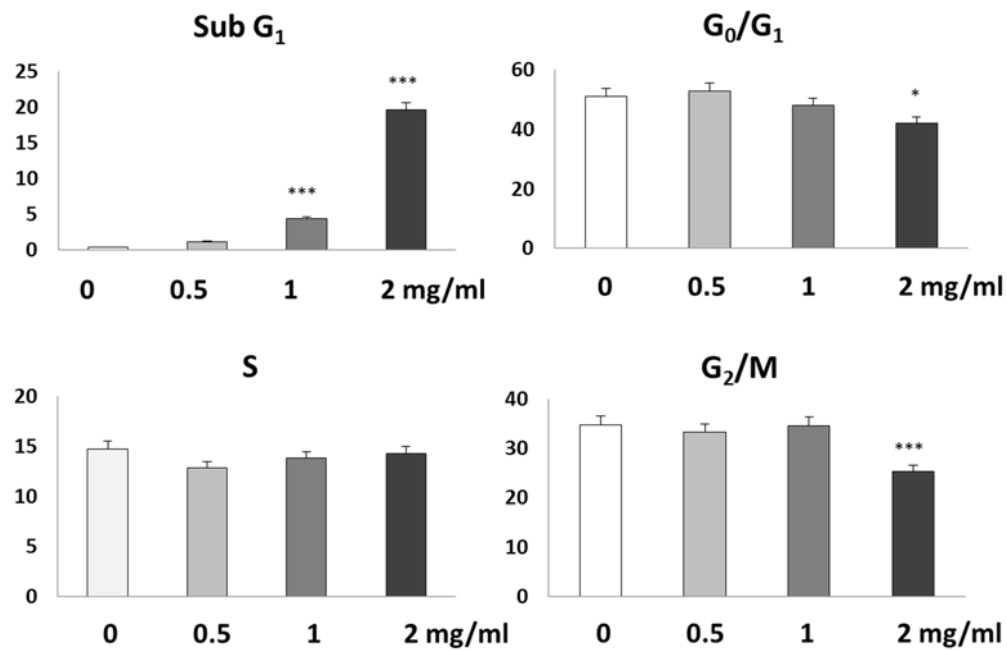

**Figure 4.** Effects of different doses of powdered MBG-75S on cell cycle phases of human Saos-2 osteoblasts after 24 hours of treatment. Statistical significance: *$p < 0.05$; *** $p < 0.005$.

The cell size and complexity of osteoblasts in the absence or the presence of different doses of MBG-75S were also evaluated by flow cytometry through FSC and 90° SSC light scatters, respectively. These properties depend on cell size, plasma membrane and intracellular organelles [33]. Significant effects of MBG-75S on these two parameters ($p < 0.005$) were observed (Figure 5), evidencing dose-dependent decreases of osteoblast size and complexity induced in by this powdered material. These effects could be due to the loss of cell anchorage produced by the powdered material and to changes of the ambient ionic concentration as consequence of ion release from the material.

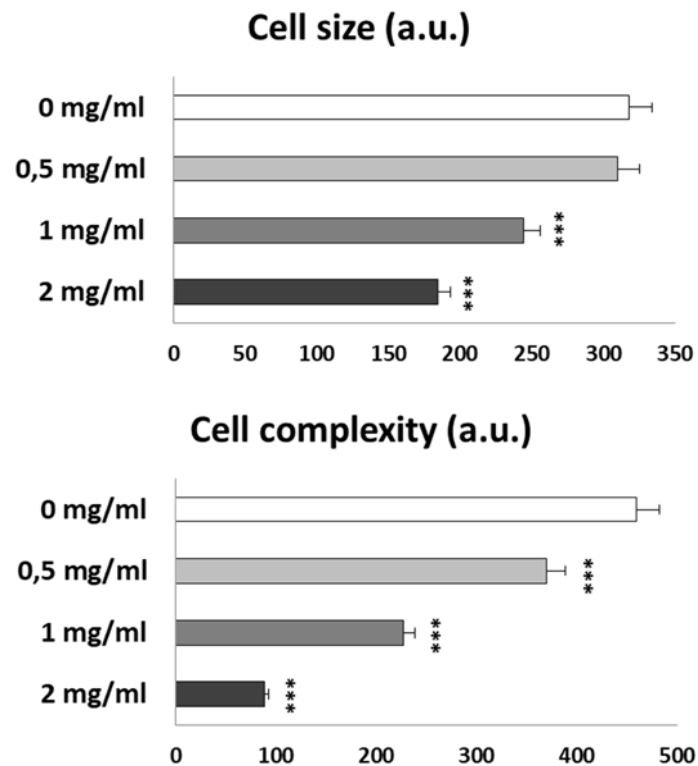

**Figure 5.** Effects of different doses of powdered MBG-75S on cell size and complexity of human Saos-2 osteoblasts after 24 hours of treatment. Statistical significance: *** $p < 0.005$.

The morphology of human Saos-2 osteoblasts in the presence of different doses of MBG-75S was observed by confocal microscopy after staining with rhodamine-phalloidin (for

F-actin filaments in red) and DAPI (for nuclei in blue). The typical characteristics of this cell type were observed in the presence of 0.5 and 1 mg/ml. However, 2 mg/ml induced cell morphology alterations in agreement with the pronounced apoptosis increase obtained by flow cytometry after treatment with this high MBG-75S dose (Figure 6). Considering the MBG-75S dose-dependent effects observed with Saos-2 osteoblasts, the dose chosen for studying the response of osteoclasts and macrophages to this material was 1 mg/ml.

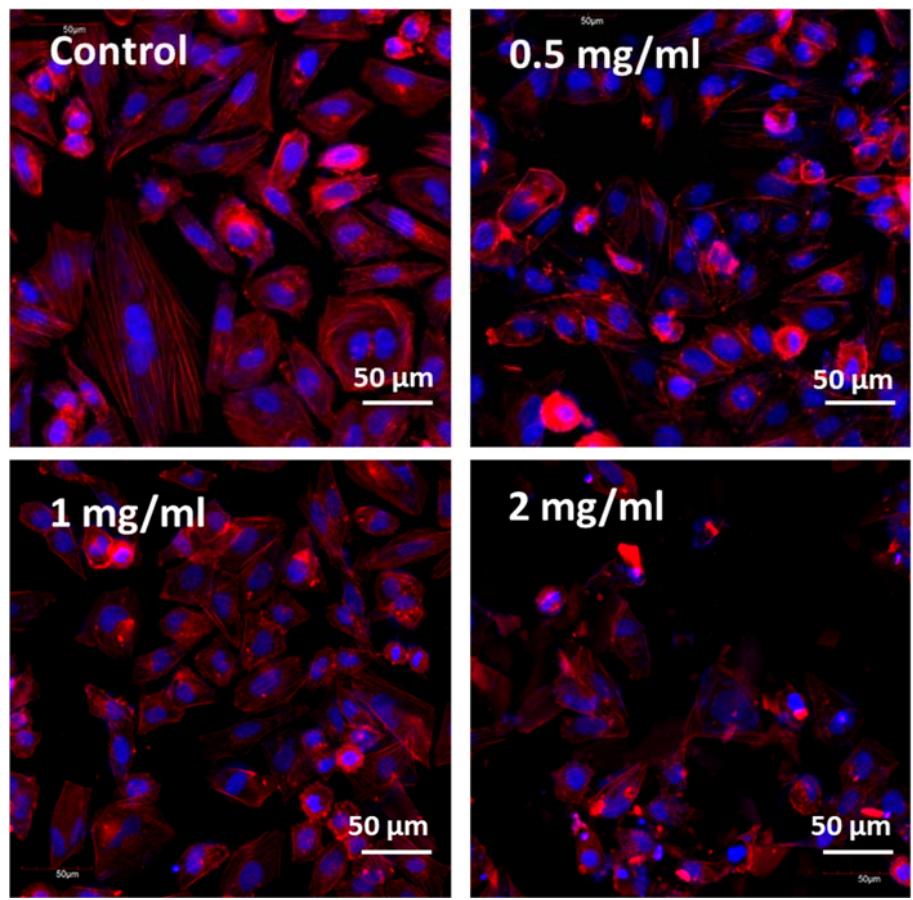

**Figure 6.** Effects of different doses of powdered MBG-75S on the morphology of human Saos-2 osteoblasts observed by confocal microscopy after 24 hours of treatment. Actin was stained with rhodamine-phalloidin (red) and cell nuclei with DAPI (blue).

*3.3. Effects of MBG-75S on osteoclast differentiation and resorption activity*

Osteoclasts can differentiate *in vitro* from macrophages by stimulation with the macrophage/monocyte colony-stimulating factor (M-CSF) and the receptor activator of nuclear factor kappa-B ligand (RANKL). RAW-264.7 is a mouse macrophage cell line retaining many of the characteristics of macrophages *in vivo* [34]. In the present study, osteoclast-like cells were obtained from RAW-264.7 macrophages after 7 days of differentiation with RANKL and M-CSF, in the absence or in the presence of 1 mg/ml MBG-75S. The morphology of these osteoclast-like cells was evaluated by confocal microscopy (Figure 7) which allowed us to observe multinucleated cells in the presence and in the absence of MBG-75S, revealing osteoclast-like cell differentiation from RAW macrophages after 7 days in both conditions. The presence of numerous actin rings, critical to define the sealing zone required for osteoclast resorption activity, was also observed (Figure 7). Two images for each group are shown in order to highlight the presence of multinucleated cells and the actin rings, that are the main characteristics of osteoclasts. A statistical analysis of the multinucleated cells has been carried out in the absence and in the presence of MBG-75S obtaining values of $10\% \pm 1\%$ of multinucleated cells in both cases in agreement with previous studies [22]. Our experiments evidence that the presence of MBG-75S does not inhibit the osteoclastogenesis, at least at the particles concentration used in this work.

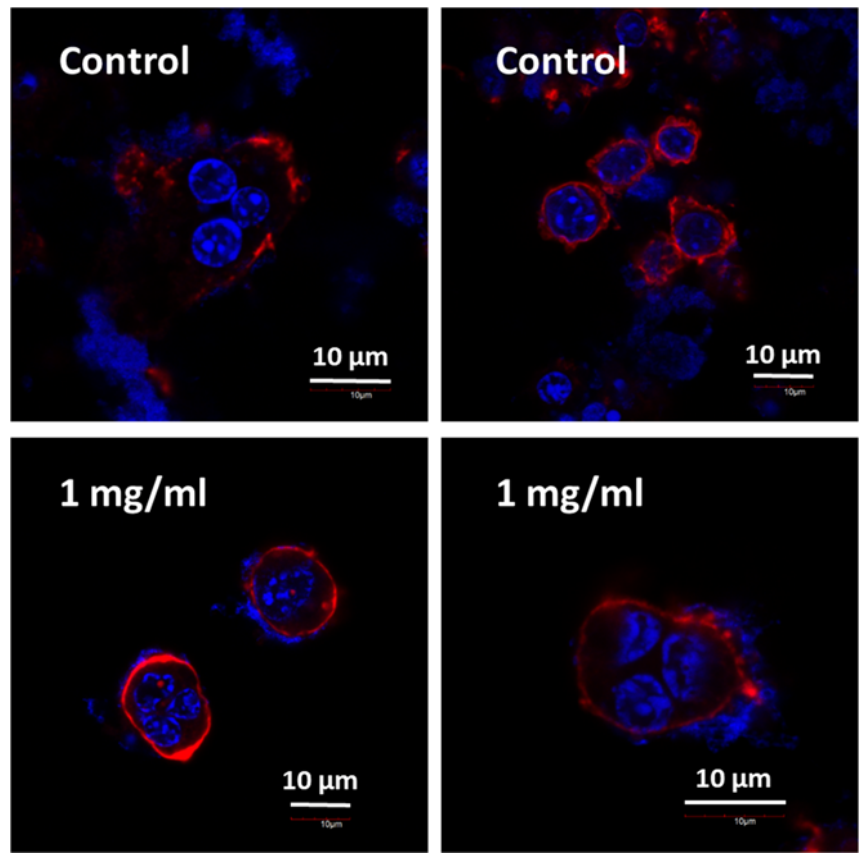

**Figure 7.** Effects of 1 mg/ml of powdered MBG-75S on the morphology of osteoclast-like cells observed by confocal microscopy after 7 days of treatment. Actin was stained with rhodamine-phalloidin (red) and cell nuclei with DAPI (blue).

The resorption cavities left by osteoclast-like cells on nanocrystalline hydroxyapatite disks after 7 days of differentiation, were evaluated by SEM. As it can be observed in Figure 8, the morphology and the number of these cavities evidenced that the resorptive activity is significantly modified by the presence of MBG-75S. Certainly, the cavities observed in both cases are of 20 to 40 micrometers in size (30 ± 10 μm), in agreement with the sizes of the actin rings previously observed (see Figure 7). However, in the absence of MBG-75S particles, osteoclasts leave marks significantly deeper with well-defined rims (Figure 8, control images) in comparison with the marks left in the presence of the MBG, where only weak dark contrasts can be observed on the surface of the nano-HA substrate (Figure 8, right images). The magnification of each image was chosen during the observation of the different samples in order to highlight the characteristics of the cavities left by osteoclasts in the presence and in the absence of MBG-75S. Higher magnifications (Figure 8, bottom images) evidence the lower resorptive activity of osteoclasts differentiated in the presence of MBG-75S. Whereas the mark left without the MBG is a well-outlined cavity, exhibiting several micrometers in depth and a rim indicating the actin sealing area, the mark left in the presence of the MBG is just a superficial erosion poorly outlined rims.

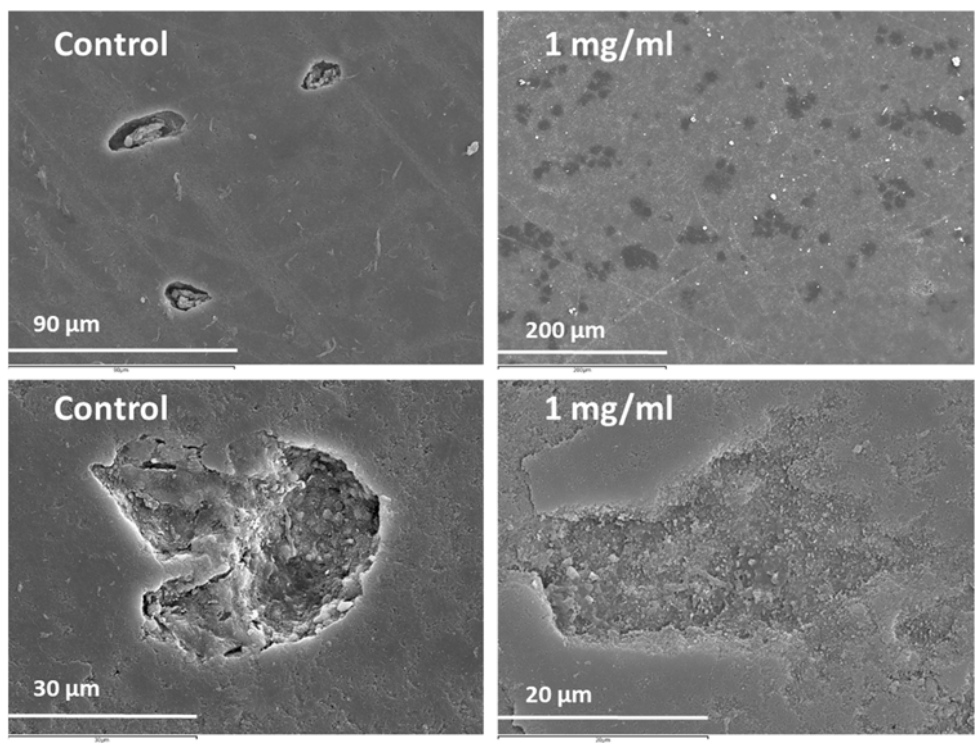

**Figure 8.** Scanning electron microscopy images of the resorption cavities left by osteoclast-like cells cultured on nanocrystalline hydroxyapatite disks after 7 days of culture in the absence (control) and the presence of 1 mg/ml of powdered MBG-75S. Bottom images shows a higher magnification of a resorption cavity in the absence (control) and the presence of MBG-75S.

To know the effects of MBG-75S on plasma membrane integrity during the differentiation process, the lactate dehydrogenase (LDH) released into the culture medium of osteoclast-like cells was measured after 3 and 7 days of treatment. Although the presence of MBG-75S produced a significant increase of LDH levels after 3 days, no significant differences between control and treated cells were observed after 7 days of differentiation, thus indicating the integrity of plasma membrane of these cells after 7 days of culture with this material.

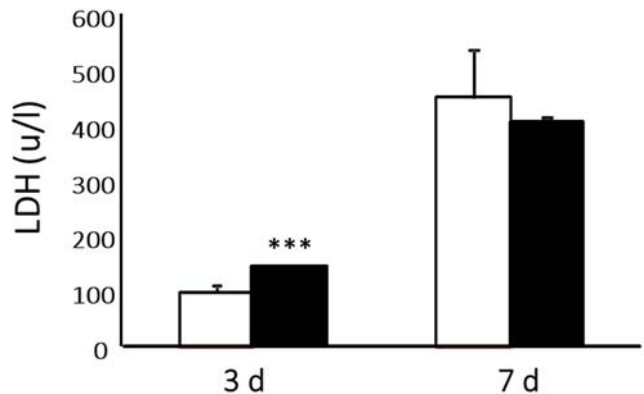

**Figure 9.** Effects of 1 mg/ml of powdered MBG-75S on lactate dehydrogenase (LDH) released into the media of osteoclast-like cells during the differentiation process after 3 and 7 days of treatment. Controls without material were carried out in parallel (white). Statistical significance: *** $p < 0.005$.

To investigate specifically the possible effect of ionic dissolution products of MBG-75S on the resorption activity of osteoclast-like cells, $Ca^{2+}$, phosphate and soluble silica species levels were measured into the culture medium during the differentiation process after 3 and 7 days of treatment with 1 mg/ml of this powdered material (Figure 10). A significant increase of $Ca^{2+}$, a significant decrease of phosphate and a very pronounced significant increase of silica levels ($p < 0.005$) were observed after 3 and 7 days of treatment with 1 mg/ml MBG-75S (Figure 10).

Different authors have demonstrated that the bone-resorbing activity of osteoclasts is regulated by extracellular $Ca^{2+}$ concentration and that high levels of this ion produce osteoclast retraction and dissipation of sealing zone, decreasing the cell spread area with actin reorganization and podosomal disassembly, which resulted in a dramatic reduction of bone resorption [35,36]. The high soluble silica levels detected into the medium can also contribute to this effect due to the inhibitory action of this ion on the osteoclastic activity [37,38]. In this sense, the ionic exchange between MBG-75S and the surrounded fluids would be facilitated by the high surface area, porosity, pore size/wall thickness ratio and, in general, by the highly ordered mesoporous structure of the MBG-75S.

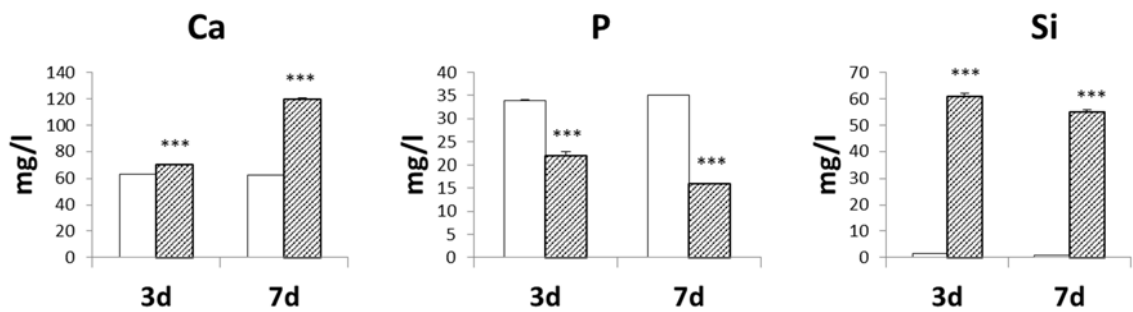

**Figure 10.** $Ca^{2+}$, phosphorous and silicon levels measured into the culture medium of osteoclast-like cells during the differentiation process after 3 and 7 days of treatment with 1 mg/ml of powdered MBG-75S. Controls without material were carried out in parallel (white). Statistical significance: *** $p < 0.005$.

Since $Ca^{2+}$ and soluble silicate ions also stimulate the proliferation and differentiation of osteoblasts [11,37,39], MBG-75S presents a high potential for bone regeneration due to its capability for releasing these two ions.

*3.4. Effects of MBG-75S on polarization of RAW-264.7 macrophages towards pro-inflammatory M1 phenotype*

The implantation of a biomaterial is accompanied by tissue injury through the surgical procedure that initiates an inflammatory response, starting with the formation of a provisional matrix. Thus, after the first contact between the biomaterial and the tissue, proteins from blood and interstitial fluids adsorb to the biomaterial surface, determining the activation of coagulation cascade, complement system, platelets and immune cells. These facts result in the formation of a transient provisional matrix and the onset of the inflammatory response. On the other hand, the immune response is additionally affected by the ions and the products eluted from the biomaterial implanted in the body. All these products could reach the bloodstream affecting lymphocytes and macrophages which release reactive oxygen species (ROS) and cytokines, events which play an important role in the inflammatory response towards the biomaterial. An immune response involves the action of all types of macrophages, classical activated macrophages (M1) in the early phase and wound-healing macrophages (M2) in the resolution stage. However, when inflammatory stimuli persist at the implant site, macrophages attached to the biomaterial can foster invasion of additional inflammatory cells by secreting chemokines like IL-8, MCP-1, MIP-1b leading to chronic inflammation [40].

In order to know if MBG-75S promotes the pro-inflammatory M1 macrophage phenotype, RAW-264.7 macrophages were cultured in the absence or the presence of 1 mg/ml of this material for 24 hours without or with *E. coli* lipopolysaccharide and interferon-γ, as inflammatory stimuli [28]. The expression of CD80 as M1 marker [29], was quantified by flow cytometry and observed by confocal microscopy with a phycoerythrin (PE) conjugated anti-mouse CD80 antibody. Previously, the effects of MBG-75S on RAW-264.7 proliferation were evaluated.

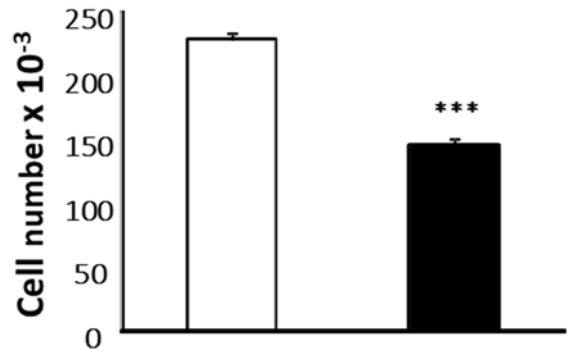

**Figure 11.** Effects of 1 mg/ml of powdered MBG-75S on proliferation of RAW-264.7 macrophages after 24 hours of treatment. Controls without material were carried out in parallel (white). Statistical significance: *** $p < 0.005$.

As it can be observed in Figure 11, MBG-75 allowed RAW-264.7 macrophages to proliferate but more slowly than control cells (white), inducing a significant decrease of the cell number ($p < 0.005$), as it has been previously observed with others powdered materials for bone tissue [41]. In previous studies with other mesoporous bioactive glass MBG-85, the release of high $Ca^{2+}$ concentrations reduced Saos-2 osteoblast proliferation whereas Si concentration did not produce negative effect on the cell proliferation [42]. The effects of 1 mg/ml of powdered MBG-75S on pro-inflammatory M1 macrophage phenotype were quantified by flow cytometry through the CD80 expression as described above in both basal and LPS/IFN-γ stimulated conditions. As it can be observed in Figure 12, the addition of LPS/IFN-γ induced a significant increase of $CD80^+$ macrophages ($p < 0.005$) in the absence (white) and in the presence (black) of MBG-75S. However, no significant differences were observed between control cells and MBG-75S treated macrophages in both basal and LPS/IFN-γ stimulated conditions. Figure 13 shows CD80 expression of M1 RAW-264.7 macrophages observed by confocal microscopy after 24 hours of treatment with *E. coli* lipopolysaccharide and interferon-γ as inflammatory stimuli.

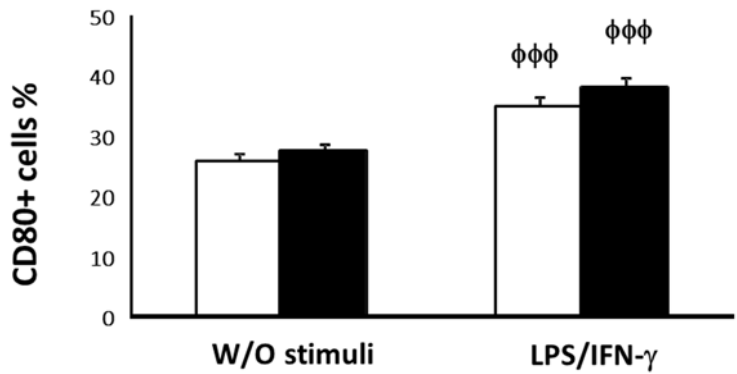

**Figure 12.** Effects of 1 mg/ml of powdered MBG-75S on pro-inflammatory M1 macrophage phenotype after 24 hours of treatment without or with *E. coli* lipopolysaccharide and interferon-γ (LPS/IFN-γ) as inflammatory stimuli. Controls without material were carried out in parallel (white). Statistical significance: $^{\phi\phi\phi}$ $p < 0.005$ (comparison between basal and LPS/IFN-γ stimulated conditions).

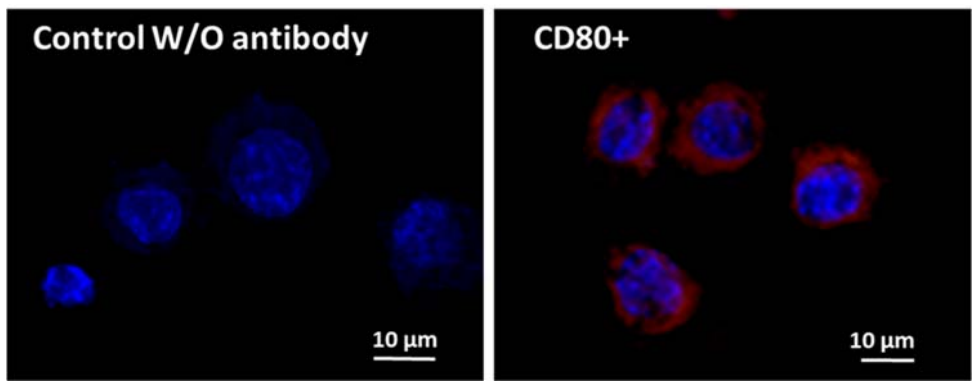

**Figure 13.** CD80 expression of M1 RAW-264.7 macrophages observed by confocal microscopy after 24 hours of treatment with *E. coli* lipopolysaccharide and interferon-γ as inflammatory stimuli. CD80 (red) was detected with PE conjugated anti-mouse CD80 antibody and nuclei were stained with DAPI (blue). Controls without antibody were carried out in parallel (left).

These results evidence that MBG-75S did not induce the macrophage polarization towards M1 pro-inflammatory phenotype, ensuring an appropriate immune response to this mesoporous bioactive glass.

Bone remodeling is a very complex process that involves different cellular types to reach an equilibrium between bone formation, bone resorption and inflammatory response. In a scenario of degenerative bone disease as osteoporosis, bone formation by osteoblasts is decreased respect to resorption driven by osteoclasts. For this reason, the osteoinductive effect of MBGs has been widely studied by different research groups, demonstrating significant increases in osteoblast proliferation [43-45] and a key role as differentiation stimuli from mesenchymal stem cells to osteoblast phenotype [46,47]. However, the effect of MBGs over other cell types as osteoclasts and macrophages have been poorly studied.

Bone grafts intended for bone regeneration purposes in osteoporotic patients not only should boost the osteoblastic pathway, but also to decrease the resorptive activity of osteoclast without inhibiting osteoclastogenic function. The inhibition of osteoclastogenesis could result in adynamic bone scenarios, similarly to those observed with the systemic administration of bisphosphonates, thus preventing bone regeneration.

Besides, the bone graft should not produce a strong inflammatory response, without inhibiting the innate immune response mediated by macrophages. This response is mandatory to trigger the tissue healing process, but it must be controlled to avoid chronic inflammations that would impede the appropriated bone healing.

The results observed in this study with MBG-75S with the three cell types (osteoblasts, osteoclasts and macrophages) point out very interesting responses to be considered as bone grafts, especially for osteoporotic patients. For concentration of 1 mg/ml, MBG-75S does not alter the osteoblasts cycle and their morphology. Besides, osteoclastogenesis from macrophages was not hindered in the presence of this material and their plasma membrane were not altered, indicating that the osteoclast formation pathway is not blocked by MBG-75S. However, the resorptive capability of these osteoclasts was significantly hampered, probably due to the high and fast silicon and calcium release from the materials to the culture media. In this sense, the mesoporous structure of this material is likely to play a very important role. The activity of MBG-75S over osteoclasts opens the possibility for obtaining bone grafts, which could reduce the osteoclast resorption without resulting in adynamic bone scenarios.

Finally, MBG-75S evidences an excellent behavior respect to innate immune response, as this material allows the appropriated development of macrophages without polarization towards M1 pro-inflammatory phenotype.

## 4. Conclusions

The ions released from MBGs can stimulate the expression of several genes of osteoblastic cells [16] and could also regulate immune responses by altering the ionic microenvironment between the implants and hosts [18]. In the present study, a mesoporous bioactive glass with molar composition $75SiO_2$-$20CaO$-$5P_2O_5$ (MBG-75S) has been synthetized and its effects on osteoblasts, osteoclasts and macrophages have been evaluated jointly. This MBG exhibits a high mesoporous order and large surface area and porosity, thus allowing an efficient ionic exchange of $Ca^{2+}$ and soluble silica with the surrounding media. MBG-75S shows *in vitro* biocompatibility respect to osteoblasts and concentrations up to 1 mg/ml do not alter cell cycle of Saos-2 cells. MBG-75S does not inhibit osteoclastogenesis but decreases the resorptive activity of osteoclast cells. This fact indicates that MBG-75S would maintain bone remodeling but would slow down the bone resorption, thus favoring faster bone regeneration. MBG-75S allows macrophage proliferation without inducing polarization towards M1 pro-inflammatory phenotype. This fact is indicative that MBG-75S would allow the innate immune response required for healing process without further inflammatory complications. Overall, the *in vitro* results obtained with osteoblasts, osteoclasts and macrophages suggest that MBG-

75S is an interesting candidate as bone graft for bone regeneration purposes, especially in osteoporotic patients. Further studies are currently being performed with *in vivo* models in order to determine the advantages of this MBG respect to other bioceramics; these results will be included in a future manuscript.

## Acknowledgements

M.V.R. acknowledges funding from the European Research Council (Advanced Grant VERDI; ERC-2015-AdG Proposal No.694160). N.G.C. is greatly indebted to Ministerio de Ciencia e Innovación for her predoctoral fellowship. L.C. acknowledges the financial support from Comunidad de Madrid (Spain, CT4/17/CT5/17/PEJD-2016/BMD-2749). The authors also thank to Spanish MINECO (MAT2015-64831-R, MAT2016-75611-R AEI/FEDER, UE). The authors wish to thank the ICTS Centro Nacional de Microscopia Electrónica (Spain), CAI X-ray Diffraction, CAI NMR, CAI Flow Cytometry and Fluorescence Microscopy of the Universidad Complutense de Madrid (Spain) for their technical assistance.